\documentclass[useAMS,usenatbib]{mn2e}
\usepackage{graphicx}

\title[Spectral and temporal changes in 4U 1626$-$67]{Spectral and temporal changes associated with flux enhancement in 4U 1626$-$67}

\author[Chetana Jain, Biswajit Paul and Anjan Dutta]{Chetana Jain$^{1,2}$\thanks{E-mail: chetanajain11@gmail.com}, Biswajit Paul$^{2}$ and Anjan Dutta$^{1}$\\
$^{1}$Department of Physics and Astrophysics, University of Delhi,  Delhi 110007, India\\
$^{2}$Raman Research Institute, Sadashivnagar, C. V. Raman Avenue, Bangalore 560080, India}
\begin{document}

\pagerange{\pageref{firstpage}--\pageref{lastpage}} \pubyear{2009}
\maketitle
\label{firstpage}

\begin{abstract}

4U 1626$-$67 is an accretion powered X-ray pulsar that shows remarkably stable X-ray luminosity above 
hours timescale and gradual intensity variation on a few years timescale. Unlike the other 
high magnetic field binary X-ray pulsars, the spin-up or spin-down rate of the neutron star 
is also very stable upto several years. Here we report a significant increase in the X-ray intensity 
in the long term \emph{RXTE}-ASM light curve of 4U 1626$-$67. Similar enhancement in the X-ray 
flux has also been detected in the \emph{Swift}-BAT light curve. The increase in the X-ray 
flux took place over a long period of about 100 days and there appears to be two episodes of 
flux enhancement. We have investigated the spectral and timing features of 4U 1626$-$67 
during its current state of enhanced flux emission with data obtained from the 
Proportional Counter Array and the High-Energy X-ray Timing Explorer on board 
the \emph{Rossi X-ray Timing Explorer}. We report the detection of a torque reversal 
to spin-up in 4U 1626-67. The source has entered a new spin-up phase with a spin-up rate 
of 4.02(5) $\times$ 10$^{-13}$ Hz s$^{-1}$. The present spin-up rate is almost half of 
the earlier spin-up and spin-down trends. A significant excess in soft X-ray photon emission 
is observed during the enhanced flux state, which is similar to the energy spectrum 
obtained during the spin-up era of the pulsar before 1990. 4U 1626$-$67 is a unique 
accretion powered X-ray pulsar in which quasi periodic oscillations have been consistently 
observed over the past $\sim$20 years. But during the recent observations, we did not 
detect a QPO at frequencies as observed in earlier observations. Instead, we report 
detection of a significant broadening in the wings of the 130 mHz peak and a change in the 
shape of the continuum of the power spectrum. These results indicate that the flux 
enhancement is not a simple case of increased mass accretion rate, but there is also a 
change in the accretion geometry in the vicinity of the neutron star. 
\end{abstract}
\begin{keywords}
X-ray: Neutron Stars - pulsars: individual (4U 1626$-$67) - X-ray: binaries
\end{keywords}

\label{Introduction}
\section{Introduction}

4U 1626$-$67 is an accretion powered X-ray pulsar which was discovered with the \emph{Uhuru} satellite 
in 1972 (Giacconi et al. 1972). The 7.7s X-ray pulsations were discovered by Rappaport et al. (1977) 
during SAS3 observations and have been extensively monitored since then. Initially the pulsar was 
found to be spinning up with a characteristic timescale of $\sim$5000 years. It underwent an abrupt 
torque reversal in 1990 and has been steadily spinning down since then (Chakrabarty et al. 1997, 
Krauss et al. 2007). Optical counterpart of the pulsar was identified as KZ TrA, a faint blue star with 
little reddening (McClintock et al. 1977; Bradt $\&$ McClintock 1983). Pulsations at 130 mHz have also 
been detected from the reprocessed optical emission (Middleditch et al. 1981). Multiple sidebands 
separated by about 0.4 mHz, are present in the power spectrum of the optical light curve, which was 
interpreted as signature of an orbital period of $\sim$42 min (Middleditch et al. 1981; Chakrabarty 
et al. 2001). But despite extensive searches, an orbital motion of the neutron star (usually observed in 
binary pulsars in the form of Doppler shift in the pulse period, or delay in the pulse arrival time) has 
never been detected in this source (Rappaport et al. 1977; Levine et al. 1988; Jain et al. 2007). It is 
believed that the neutron star most probably, has a low mass binary companion and is in a 
nearly face-on orbit. 

The X-ray spectrum of 4U 1626$-$67 was first studied in detail with the \emph{ASCA} spectrometers 
(Angelini et al. 1995) and later with instruments on board \emph{Beppo-SAX} (Owens et al. 1997; 
Orlandini et al. 1998), \emph{Chandra} (Schulz et al. 2001) and \emph{XMM$-$Newton} 
(Krauss et al. 2007). The X-ray continuum spectrum is evidently correlated with the torque 
state. During the 1977$-$1990 spin-up phase, the spectrum was well described by an absorbed 
blackbody, a power law and a high energy cutoff (Pravdo et al. 1979; Kii et al. 1986). But 
after the torque reversal in 1990, the time-averaged X-ray spectrum was found to be relatively 
harder (Vaughan $\&$ Kitamoto 1997, Yi $\&$ Vishniac 1999). Orlandini et al. (1998) reported 
an absorption feature at $\sim$37 keV which they attributed to be the cyclotron absorption 
line. The spectrum is rich in bright hydrogen-like and helium-like oxygen and neon emission 
lines. The blue and red Doppler shifted emission lines in the X-ray spectrum indicate that the 
lines originate in the accretion disk. However, presence of the red and blue Doppler shifted 
emission lines from the accretion disk is difficult to explain if the binary system is face-on. 
Presence of a metal rich accretion disk and absence of orbital motion led Jain et al. (2007) to suggest 
that 4U 1626$-$67 is a candidate for a neutron star with a supernova fallback accretion disk.

An important signature of the presence of an accretion disk is the detection of Quasi Periodic 
Oscillations (QPOs) in the power spectrum of the X-ray pulsar. QPOs are believed to arise due to 
inhomogeneities in the inner accretion disk. This aperiodic X-ray variability has been 
detected in several X-ray sources, such as, Cen X-3 (Takeshima et al. 1991; Raichur $\&$ Paul 2008), 
XTE J1858$+$034 (Paul $\&$ Rao 1998; Mukherjee et al. 2006), EXO 2030$+$375 (Angelini, Stella $\&$ 
Parmar 1989), XTE J0111.2$-$7317 (Kaur et al. 2007) and 3A 0535$+$262 (Finger, Wilson $\&$ Harmon 1996). 
4U 1626$-$67 is a unique X-ray pulsar 
in which QPOs have been consistently detected in the X-ray and optical observations. The QPO 
frequency has been evolving over the past 22 years (Kaur et al. 2008). The QPO central frequency 
was found to be increasing during the spin-up phase of the pulsar, whereas, a negative trend was 
observed after the torque reversal. During the spin-up phase of the source, QPOs were 
detected at a frequency of 36 mHz with the EXOSAT data (Kaur et al. 2008) and at 40 mHz 
from the \emph{Ginga} observations (Shinoda et al. 1990). During the spin-down era of the 
source, QPOs have been reported to occur at a higher frequency of about 48 mHz from 
\emph{ASCA}, \emph{Beppo}-SAX, \emph{RXTE}, and \emph{XMM-Newton} (Angelini et al. 1995; 
Owens et al. 1997; Kommers et al. 1998; Krauss et al. 2007). Large amplitude 0.3$-$1.2 mHz 
QPOs were also detected in reprocessed optical emission from both ground based and Hubble 
Space Telescope observations (Chakrabarty et al. 2001). 

Recent observations with the monitoring X-ray instruments on board the \emph{RXTE} and \emph{Swift} 
satellites, show an increase in the X-ray intensity of 4U 1626$-$67. In the present work, we have 
investigated the spectral and temporal properties of 4U 1626$-$67 during the increased flux state. 
In particular, we have searched for pulsations and the QPO feature in the power spectrum of 4U 1626$-67$ 
which was persistently present in the earlier observations. We have also studied the energy 
spectrum of the source during the current phase and have compared it with the earlier known 
spectral features. 

\section{Observations and analysis}

4U 1626$-$67 is a medium intensity persistent X-ray source among the few hundred bright X-ray 
sources regularly monitored by the All Sky Monitor (ASM) on board \emph{Rossi X-ray Timing Explorer} 
(\emph{RXTE}). Figure 1 (thin line) shows the long term 1.5$-$12 keV ASM light curve of 
4U 1626$-$67, binned with 30 d. The data used here covered the time range between MJD 50088 
to MJD 54909. The X-ray intensity was gradually decreasing until MJD 54500, when an 
increase in the X-ray flux was noticed. The inset in Figure 1 shows the expanded view near 
the onset of the flux enhancement. There seems to be two episodes of X-ray flux enhancement, 
both on timescales of $\sim$100 days.  

4U 1626$-$67 was also regularly monitored by the Burst Alert Telescope (BAT; Barthelmy et al. 2005) 
on board the \emph{Swift} observatory (Gehrels et al. 2004). The long term 15$-$50 keV \emph{Swift}-BAT 
light curve, binned with 30 d is also shown in Figure 1 (thick line) alongwith the ASM light curve. The 
observations covered the time range from MJD 53414 to MJD 54920. A sudden increase in 
the X-ray flux was first reported by Krimm et al. (2008). The inset in Figure 1 (thick line) 
shows the expanded view (in the same units) near the onset of the flux enhancement. The flux 
change on timescales of $\sim$100 days is also clear in the \emph{Swift}-BAT light curve. 

\begin{figure}
\centering
\includegraphics[height=3.5in, width=3.0in, angle=-90]{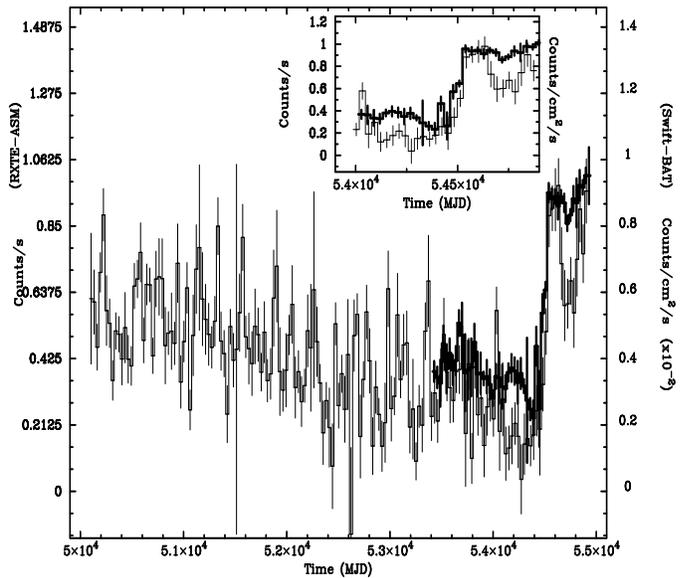}
\caption{\emph{RXTE}-ASM and \emph{Swift}-BAT light curves of 4U 1626$-$67, binned with 30 d. 
The 15$-$50 keV \emph{Swift}-BAT light curve is shown with a thick line over the 1.5$-$12 keV 
\emph{RXTE}-ASM light curve (thin line). The inset figure shows the expanded view near the 
onset of the flux enhancement which occured around MJD 54500 (thick line is 
the \emph{Swift}-BAT light curve in the same units as in the main panel).}
\end{figure}

We have also analysed data obtained from the Proportional Counter Array (PCA) 
and the High Energy X-ray Timing Explorer (HEXTE) on board the \emph{RXTE}. 
The PCA consists of five xenon/methane proportional counter units (PCUs) and is sensitive in the 
energy range of 2$-$60 keV with an effective area of 1300$-$6500 cm$^{2}$, depending on the 
number of PCUs ON (Jahoda et al. 1996). The HEXTE operates in the energy range 15$-$250 keV 
and consists of two clusters of phoswich scintillation detectors, each having a 
collecting area of 800 cm$^{2}$ (Gruber et al. 1996).

Table 1 gives a log of the \emph{RXTE} observations used for the 
present analysis. We have studied the temporal and spectral variations in 4U 1626$-$67, using 
data collected in the Standard-1 and the Standard-2 mode of the \emph{RXTE}-PCA, respectively. 
The archive mode data of the HEXTE was used for the spectral analysis. 

\begin{table}
\caption{Log of \emph{RXTE} observations analyzed in the present work}
\begin{centering}
\begin{tabular}{ | c | c | c | c |} 
\hline\hline
Year	& Observation ID	& Number of  	& Total Exposure \\
	& 			& pointings	& (ks)\\
\hline
1996	& P10101		&	3	&	153\\
	& P10144		&	1	&	10\\
	& P20146		&	2	&	2\\
1997	& P20146		&	12	&	7\\
1998	& P30058		&	2	&	40\\
	& P30060		&	7	&	53\\
2008	& P93431		&	2	&	7\\
2009	& P94423		&	3	&	9\\
\hline
\hline
\end{tabular}
\end{centering}
\label{Table 1 }  
\end{table}

\subsection{Timing Analysis}
We searched for pulsations using data obtained in the Standard 1 mode of the 
\emph{RXTE}-PCA. The background counts were simulated using the 
\textsc{ftool} - \textsc{runpcabackest} and subtracted from the source light 
curve. The photon arrival times were converted to the solar system barycenter 
using the \textsc{ftool} - \textsc{faxbary}. We searched for the spin period of the neutron 
star using the \textsc{efsearch} tool of the HEAsoft analysis 
package, \textsc{ftools} ver 6.5.1. This tool folds the light curve with a large number of trial 
periods around an approximate period and determines the best period by the $\chi^{2}$ maximization 
technique. Using this method, we determined a pulse period of 7.67941(1) s 
at MJD 54530.4, 7.67945(7) s at MJD 54538.1 and 7.67848(2) at MJD 54984.5 (Jain $\&$ Paul 2009). 
This gives a spin-up rate of 4.02(5) $\times$ 10$^{-13}$ Hz s$^{-1}$. Similar behaviour based on the 
Fermi/GBM data has also been reported by Camero-Arranz et al. (2009).  

Figure 2 shows the normalized pulse profile of 4U 1626$-$67 binned into 64 phasebins. 
The top panel shows the pulse profile generated from data obtained from \emph{RXTE}-PCA 
observations made in 1996 (ObsID 10101-01-01-00) and folded with a period of 7.66735 s 
(Jain et al. 2007). The bottom panel shows the normalized pulse profile of 2008 
data (ObsID 93431-01-01-00) folded with a period of 7.67941 s. Although the 
two profiles resemble the characteristic bi-horned profile of the 
source (Levine et al. 1988), the pulse fraction is significantly different in the 
two cases. Moreover, the peaks in the 2008 pulse profile are significantly sharper 
with a higher amplitude variation than those in the 1996 pulse profile.

\begin{figure}
\centering
\includegraphics[height=3.5in, width=4.0in, angle=-90]{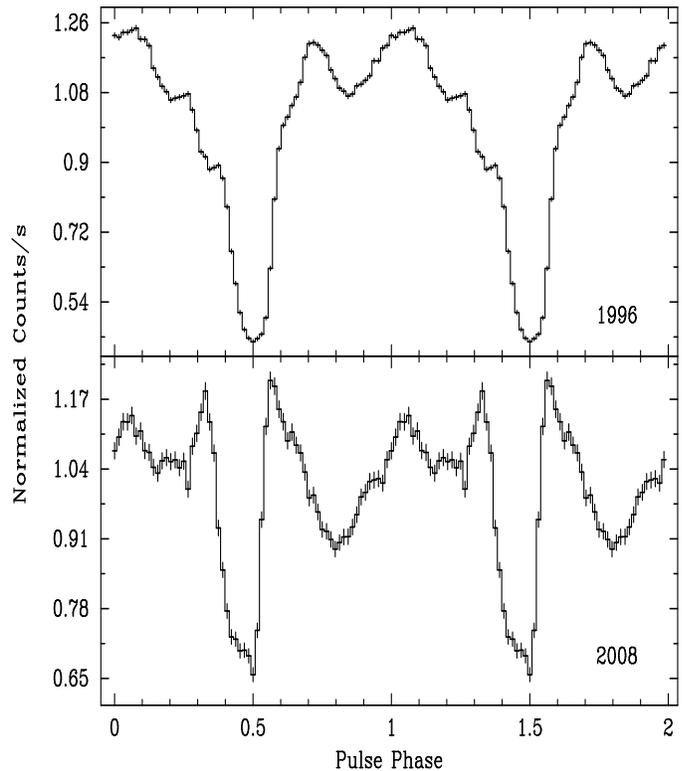}
\caption{The 2-60 keV normalized pulse profile of 4U 1626$-$67 generated from 
observations made with \emph{RXTE}-PCA during 1996 (ObsID 10101-01-01-00) and 
2008 (ObsID 93431-01-01-00). The background subtracted light curves were folded into 64 phasebins 
with a period of 7.66735 s and 7.67941 s, respectively.}
\end{figure}

We searched for QPOs in the 2$-$60 keV energy band, using the data collected in the Standard 1 mode 
of PCA, which has a time resolution of 0.125 s. We created the power density spectrum (PDS) using 
the \textsc{ftool} - \textsc{powspec} for all the sufficiently long \emph{RXTE} observations 
(See Table 1). The PDS were normalized such that their integral gives the squared rms 
fractional variability and the white noise level was subtracted. Figure 3 shows some representative power spectra generated from some of the observations mentioned in Table 1. The dates of observation are mentioned in each panel above their respective observation IDs. 

\begin{figure*}
\centering
\includegraphics[height=7.0in, width=6.5in, angle=-90]{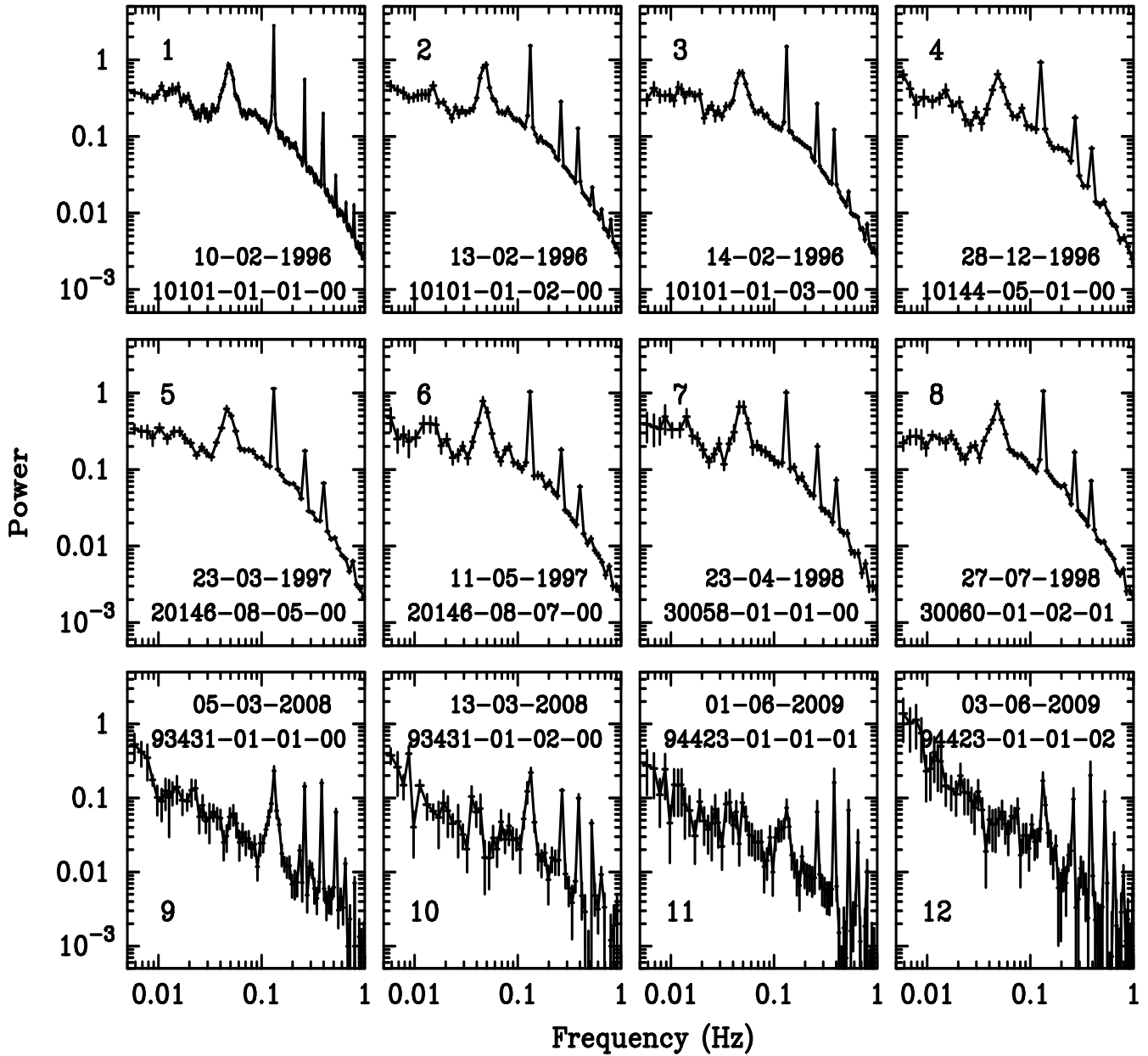}
\caption{Some representative X-ray power spectra of 4U 1626$-$67 generated from the \emph{RXTE}-PCA observations. The date of observation is mentioned in each panel above the respective observation ID.} 
\end{figure*}

\begin{figure} 
\centering
\includegraphics[height=3in, width=2.5in, angle=-90]{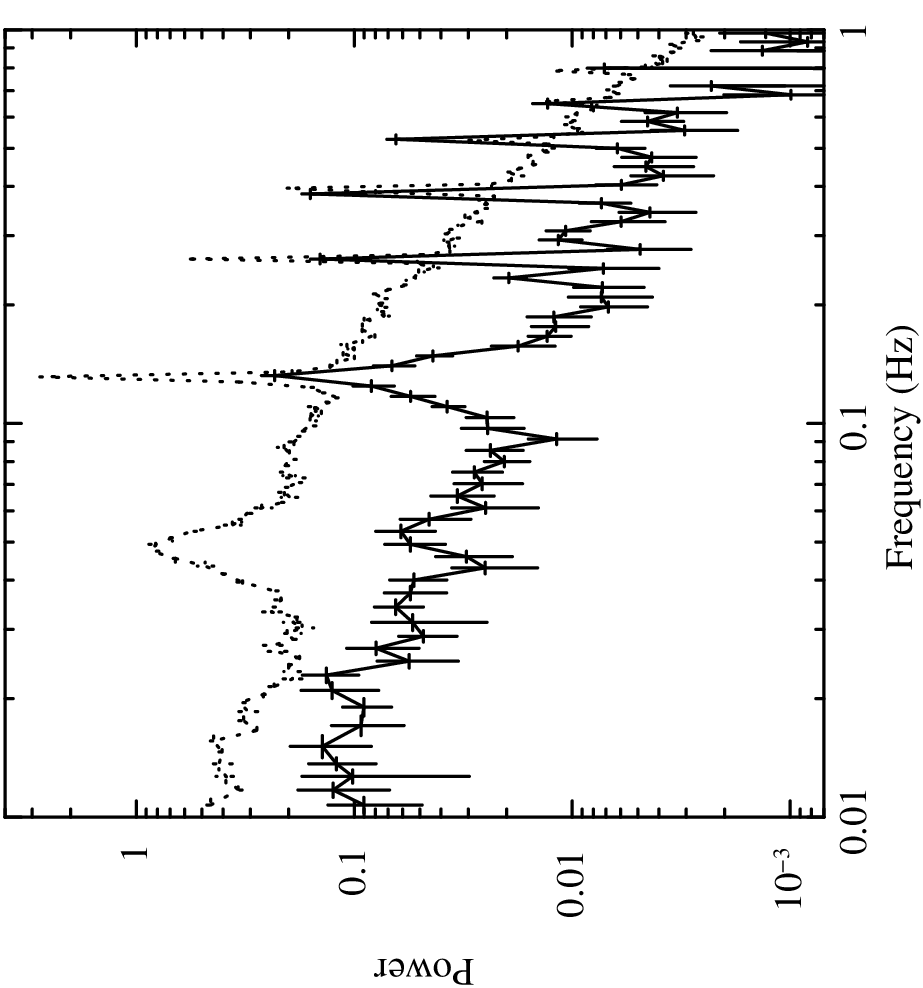}
\caption{X-ray power spectra of 4U 1626$-$67 from the \emph{RXTE} observations of 1996 and 2008. The 
upper, dotted curve is the X-ray spectrum of 1996 observation (ObsID 10101-01-01-00; Figure 3-panel 1) 
and the lower, solid curve is from 2008 observation (ObsID 93431-01-01-00; Figure 3-panel 9).}
\end{figure}

The narrow main peak at around 0.130 Hz corresponds to the spin frequency of the neutron star. Multiple 
harmonics are also seen in the PDS of the source. In addition to the main peak, a QPO feature is also 
seen at $\sim$48 mHz. However, the QPO feature is not present in the PDS generated from observations 
made in 2008 and 2009 (Figure 3 (9), (10), (11), (12)). In all the observations prior 
to 2008, the power spectrum consists of a broken power-law. As opposed to this, there is a 
remarkable difference in the power spectra generated from 2008 and 2009 observations. 

Figure 4 shows the power spectrum generated from the 2008 observations 
(ObsID 93431-01-01-00; Figure 3-panel 9) overlaid with a typical power spectrum 
from a 1996 observation (ObsID 10101-01-01-00; Figure 3-panel 1). In the 2008 power spectrum, 
absence of a QPO at $\sim$48 mHz is clearly seen. One more feature that stands out is the 
substantial broadening of the wings of the pulse peak in the power spectrum during the 2008 observation. 
The r.m.s variability in the low frequency noise component of the power spectrum was 14.6\%, 
whereas, the r.m.s variability in the broad wings of the fundamental pulse peak and the fundamental 
pulse peak itself was 6.9\% and 5.2\%.
This broad feature at $\sim$ 130 mHz was never detected before and hence provides the first 
instance of a strong coupling between the aperiodic and periodic X-ray variability in the source.  
 
\subsection{Spectral Analysis}
The pulse-phase averaged spectra of 4U 1626$-$67 were generated from the data collected in the 
Standard 2 mode of the \emph{RXTE}-PCA and the archive mode of the \emph{RXTE}-HEXTE. 
The background counts for the PCA data were simulated using the \textsc{ftool} - 
\textsc{runpcabackest} and 
subtracted from the source spectrum. In case of the HEXTE data, the source and the 
background spectra were generated using the \textsc{ftool} - \textsc{hxtback}. These were 
then corrected for the detector dead time using the \textsc{ftool} - \textsc{hxtdead}. 
In order to bring out the significant evolution in the X-ray spectrum, we have generated 
the spectrum from one of the 1996 observation (ObsID 10101$-$01$-$01$-$00) and the 2008 
observation (ObsID 93431$-$01$-$01$-$00). 

Figure 5 shows the 3$-$20 keV and 15$-$50 keV energy spectrum of the source from the 
1996 \emph{RXTE}-PCA and \emph{RXTE}-HEXTE observations. A simultaneous fit of the two 
spectrum consists of an intrinsic absorption, a powerlaw, a gaussian, a high energy cutoff 
and cyclotron resonance feature (CRF). The bottom panel of Figure 5 shows the residual of 
the fit. The best fit spectral parameters (with 1\% systematic errors) are given 
in Table 2. The spectrum is similar to the broadband continuum spectrum obtained from 
the \emph{Beppo}-SAX observations made after the torque reversal in 1990 
(Owens et al. 1997; Orlandini et al. 1998). The \emph{Beppo}-SAX energy spectrum was 
well fit by a model consisting of a low energy absorption, a blackbody, a powerlaw 
and a high energy cutoff. However, the best fit blackbody temperature was $\sim$0.3 keV which 
is much lower than the sensitivity of the \emph{RXTE}-PCA detector. We have obtained a high 
energy cutoff of 6.45 keV which is also quite small as compared to $\sim$20 keV obtained 
from the \emph{Beppo}-SAX observations.

\begin{figure}
\centering
\includegraphics[height=3in, angle=-90]{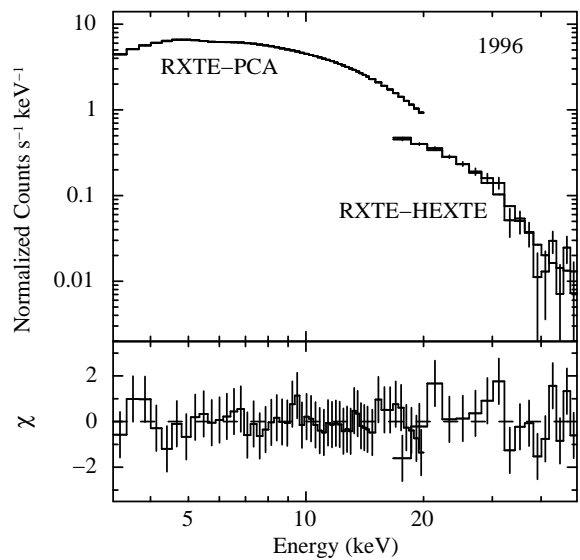}
\caption{The X-ray spectrum of the 1996 \emph{RXTE}-PCA and \emph{RXTE}-HEXTE observations of 4U 1626$-$67. The 
model components are intrinsic absorption, powerlaw, a gaussian and a high energy cutoff. The 
bottom panel shows the residual of the fit.}
\end{figure}

The 3$-$20 keV \emph{RXTE}-PCA and the 15$-$50 keV \emph{RXTE}-HEXTE energy 
spectrum obtained from the 2008 observations are shown in Figure 6. The spectra were 
simultaneously fitted with a model consisting of a powerlaw, a blackbody, a gaussian, high 
energy cutoff and cyclotron resonance feature (CRF). In Table 2, we have given the best fit 
spectral parameters (with 1 \% systematic errors). The observations made with \emph{HEAO-1} 
(Pravdo et al. 1979) and X-ray astronomy satellite \emph{Tenma} (Kii et al. 1986) during 
the earlier spin-up phase of 4U 1626$-$67 also showed similar energy spectrum. 

\begin{figure}
\centering
\includegraphics[height=3in, angle=-90]{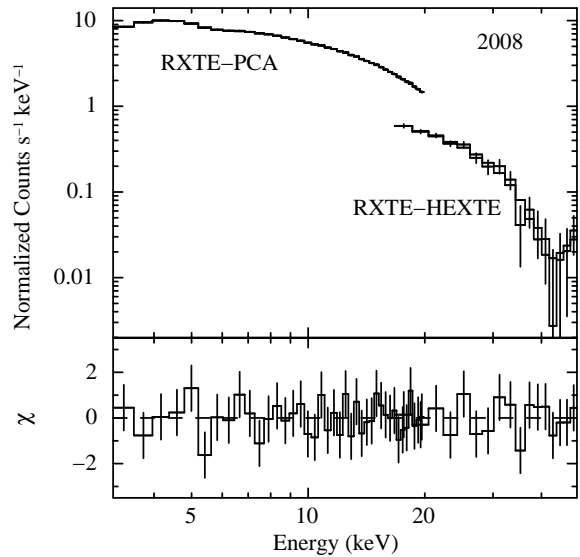}
\caption{Energy spectrum of 4U 1626$-$67 obtained from the 2008 \emph{RXTE}-PCA observations. The model 
consists of a blackbody, a powerlaw and a gaussian component.}
\end{figure} 

\begin{table}
\caption{Best fit spectral parameters of 4U 1626$-$67, obtained from the 
\emph{RXTE}-PCA and \emph{RXTE}-HEXTE observations.}
\begin{tabular}{|l|l|l|}
\hline\hline
\multicolumn{1}{|l|}{Parameter}	& \multicolumn{2}{|l|}{Value} \\
\hline
				& \multicolumn{1}{|l|}{1996}	& \multicolumn{1}{|l|}{2008}\\
\hline
N$_{H}$ (10$^{22}$ cm$^{-2}$)	&    	0.65		&	0.82	\\
Photon index ($\Gamma$)		&    	0.59(4)		&	0.74(7)	\\
norm (photons keV$^{-1}$ cm$^{-2}$ s$^{-1}$)		&    	0.0079(6)	&	0.012(2)\\	
Gaussian line (keV)		&  	6.56(40)	&	6.33(39)	\\
Gaussian width (keV)		&	1.15(21)	&	1.43(38)	\\
Gaussian norm (photons cm$^{-2}$ s$^{-1}$)			&	0.00054(32)	&	0.0019(18)	\\
Blackbody, kT (keV)		&	-		&	0.66(12)		\\
Blackbody, norm (km$^{2}$ kpc$^{-2}$) & - &       44(7)\\
cutoff E (keV) 			& 	6.45775(95)	&	17.7(8)	\\
foldE (keV)   			&     	32(5)		&	26(12)	\\
Energy$_{(CRF)}$ (keV)		&	39.5(1.8)	&	40.3(1.2)	\\
Width$_{(CRF)}$ (keV) 		&	13.8(3.0)   	&	4.8(4.5)	\\
Depth$_{(CRF)}$			&	2.02(28)	&	2.9(1.7)	\\
Reduced $\chi^{2}$ (d.o.f.)	&     	0.59 (57)	&	0.61 (43)	\\
\hline
\end{tabular}
\label{Table 2 }  
\end{table}

Figure 7 shows the evolution of the 3$-$50 keV X-ray spectrum from 1996 observation to 
the 2008 observation. In the figure, the best fit spectrum of the 1996 observation 
(dotted curve) is shown with the spectrum generated from the 2008 observations. The bottom 
panel in the figure gives the ratio between the two spectra. It is clear that there is an 
excess soft-photon emission in the spectrum generated from the 2008 observations. The powerlaw 
photon index has also increased during the recent observation.

\begin{figure}
\centering
\includegraphics[height=3in, angle=-90]{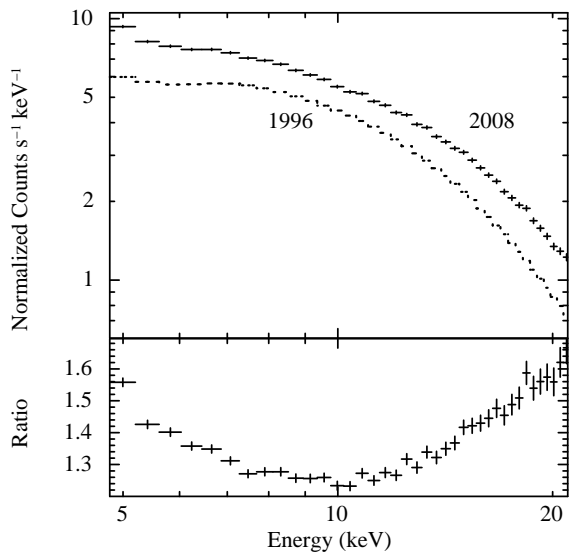}
\caption{The best fit spectrum of 4U 1626$-$67, generated from the 1996 (dotted line) and 
2008 (solid line) \emph{RXTE} observations. The bottom panel shows the ratio of the two spectra.}
\end{figure} 

\section{Discussion}

4U 1626$-$67 is a unique system in which the X-ray flux has been gradually 
decreasing over the past $\sim$30 years. Since the discovery of X-ray pulsations in 1977, 
the source has shown a smooth spin evolution history. Initially, it was found to be spinning 
up with a characteristic timescale of 5000 yr, but after an abrupt torque reversal, 
the neutron star started spinning down at about the same rate (Chakrabarty et al. 1997; 
Krauss et al. 2007). The recent flux enhancement has opened up many questions regarding 
the stability of the accretion disk. We have found a sudden increase in the X-ray flux 
of 4U 1626$-$67, which could have also caused significant changes in the geometry of the 
accretion column.  

The power spectra of X-ray pulsars consist of narrow peaks originating from the periodic signal and 
its harmonic lines which are often accompanied with aperiodic features like broad bumps, wiggles and 
steep low frequency excess (Wijnands $\&$ van der Klis 1999). The periodic variabilites arise due to the rotation induced motion of the accretion column, through which the infalling matter is funneled towards the magnetic poles of the neutron star. Whereas, any instability in the emissivity of the accretion column, can give rise to an aperiodic variability (Burderi et al. 1997). If the inhomogeneities occur far away from the surface of the neutron star, then the periodic and aperiodic variations 
occur independent of each other (Angelini et al. 1989). But if the aperiodic variability 
occurs near the accretion column, then they are modulated at the pulsar spin frequency (Makishima et al. 1988). Lazzati $\&$ Stella (1997) found a 
significant coupling between the aperiodic and periodic variabilities in Vela X-1 and 4U 1145$-$62. A 
significant coupling was also reported in the accretion powered millisecond binary 
pulsar SAX J1808.4$-$3658 (Menna et al. 2003), Her X-1 (Moon $\&$ Eikenberry 2001) and 
Cen X-3 (Raichur $\&$ Paul 2008). The coupling between the aperiodic and periodic variations is 
expected to alter the shape of the power spectrum because of the convolution of the two signals 
(Burderi, Robba $\&$ Cusumano 1993). The coupling results in a broadening of the base of the power 
spectrum peaks, as if, the peaks are riding on a QPO.  

X-ray spectrum and rapid X-ray variability of accreting compact objects have a common origin and are 
therefore expected to be correlated. In the case of 4U 1626$-$67, Yi $\&$ Vishniac (1999) 
had found that the torque reversal in 1990 was accompanied by spectral transition and a change in the 
luminosity. During the spin-up phase of 4U 1626$-$67, the phase averaged spectrum was well fit by a 
blackbody temperature \emph{k}T of $\sim$0.6 and a powerlaw photon index of $\sim$1 (Pravdo et al. 
1979; Kii et al. 1986). Vaughan $\&$ Kitamoto (1997) found that the spin transition had a significant 
effect on the 2$-$10 keV energy spectrum. The powerlaw photon index changed from 1.6 (Pravdo et al. 
1979) to $\sim$0.6 (Owens et al. 1997). The blackbody temperature also decreased from $\sim$0.6 keV 
to $\sim$0.3 keV. We have found similar spectral transition in the present high flux state of the 
source. The spectral parameters derived during the present high flux state are similar to those 
determined during the earlier spin-up phase. 

\section*{Acknowledgments}
We thank the anonymous referee for some useful comments on improving this paper. This research 
has made use of the data obtained through the High Energy Astrophysics Science Archive 
Research Center Online Service, provided by the NASA/Goddard Space Flight Center. In particular, we 
thank the \emph{RXTE}-ASM teams at MIT for provision of the ASM data.

\label{lastpage}

\begin{thebibliography}{}
\bibitem{}
Angelini L., Stella L., Parmar A. N., 1989, ApJ, 346, 906
\bibitem{}
Angelini L., White N. E., Nagase F., Kallman T. R., Yoshida A., Takeshima T., Becker C., Paerels F., 1995 ApJL, 449, 41
\bibitem{}
Barthelmy S. D., Barbier L. M., Cummings J. R. et al., 2005, Space Science Reviews, 120, 143
\bibitem{}
Bradt H. V. D., McClintock J. E., 1983, ARA$\&$A, 21, 13
\bibitem{}
Burderi L., Robba N. R., Cusumano G., 1993, Advances in Space Research, 13, 291
\bibitem{}
Burderi L., Robba N. R., La Barbera N., Guainazzi M., 1997, ApJ, 481, 943
\bibitem{}
Camero Arranz A., Finger M. H., Ikhsanov N. R., Wilson-Hodge C. A., Beklen E., 2009, arXiv0906.4224
\bibitem{}
Chakrabarty D., Bildsten L., Grunsfeld J. M. et al., 1997, ApJ, 474, 414
\bibitem{}
Chakrabarty D., Homer L., Charles P. A., O'Donoghue D., 2001, ApJ, 562, 985
\bibitem{}
Finger M. H., Wilson R. B., Harmon B. A., 1996, ApJ, 459, 288
\bibitem{}
Gehrels N., Chincarini G., Giommi P. et al., 2004, ApJ, 611, 1005
\bibitem{}
Giacconi R., Murray S., Gursky H., Kellogg E., Schreier E., Tananbaum H., 1972, ApJ, 178, 281
\bibitem{}
Gruber D. E., Blanco P. R., Heindl W. A., Pelling M. R., Rothschild R. E., Hink P. L., 1996, A\&AS, 120, 641
\bibitem{}
Jahoda K., Swank J. H., Giles A. B., Stark M. J., Strohmayer T., Zhang W., Morgan E. H., 1996, SPIE, 2808, 59
\bibitem{}
Jain C., Paul B., Joshi K., Dutta A., Raichur H., 2007, JA$\&$A, 28, 175
\bibitem{}
Jain C., Paul B., 2009, ATel 2095
\bibitem{}
Kaur R., Paul B., Raichur H., Sagar R., 2007, ApJ, 660, 1409
\bibitem{}
Kaur R., Paul B., Kumar B., Sagar, R., 2008, ApJ, 676, 1184
\bibitem{}
Kii T., Hayakawa S., Nagase F., Ikegami T., Kawai N., 1986, PASJ, 38, 751
\bibitem{}
Kommers J. M., Chakrabarty D., Lewin W. H. G., 1998, ApJL, 497, 33
\bibitem{}
Krauss M. I., Schulz N. S., Chakrabarty D., Juett A. M., Cottam J. 2007, ApJ, 660, 605
\bibitem{}
Krimm H. A., Barthelmy S. D., Baumgartner W. et al., 2008, ATel, 1426, 1
\bibitem{}
Lazzati D., Stella L. 1997, ApJ, 476, 267
\bibitem{}
Levine A., Ma C. P., McClintock J., Rappaport S., van der Klis M., Verbunt F., 1988, ApJ, 327, 732
\bibitem{}
Makishima K., 1988, in Physics of Neutron Stars and Black Holes, ed. Y. Tanaka (Tokyo: Universal Acad.), 175
\bibitem{}
McClintock J. E., van Paradijs J., Hidajat B., Hendricks H., 1977, IAUC, 3084, 3
\bibitem{}
Menna M. T., Burderi L., Stella L., Robba N., van der Klis M., 2003, ApJ, 589, 503
\bibitem{}
Middleditch J., Mason K. O., Nelson J. E., White N. E., 1981, ApJ, 244, 1001
\bibitem{}
Moon Dae-Sik, Eikenberry S. S., 2001, ApJL, 552, 135
\bibitem{}
Mukherjee U., Bapna S., Raichur H., Paul B., Jaaffrey S. N. A., 2006, JA$\&$A, 27, 25
\bibitem{}
Orlandini M., Fiume D. D., Frontera F. et al., 1998, ApJL, 500, 163
\bibitem{}
Owens A., Oosterbroek T., Parmar A. N., 1997, A$\&$A, 324, L9
\bibitem{}
Paul B., Rao A. R., 1998, A$\&$A, 337, 815
\bibitem{}
Pravdo S. H., White N. E., Boldt E. A. et al., 1979, ApJ, 231, 912
\bibitem{}
Raichur H., Paul B., 2008, ApJ, 685, 1109
\bibitem{}
Rappaport S., Markert T., Li F. K., Clark G. W., Jernigan J. G., McClintock J. E., 1977, ApJL, 217, 29
\bibitem{}
Schulz N. S., Chakrabarty D., Marshall H. L., Canizares C. R., Lee J. C., Houck J., 2001, ApJ, 563, 941
\bibitem{}
Shinoda K., Kii T., Mitsuda K., Nagase F., Tanaka Y., Makishima K., Shibazaki N., 1990, PASJ, 42, L27
\bibitem{}
Takeshima T., Dotani T., Mitsuda K., Nagase F., 1991, PASJ, 43, L43
\bibitem{}
Vaughan B. A., Kitamoto S., 1997, astro-ph/9707105
\bibitem{}
Wijnands R., van der Klis M., 1999, ApJ, 514, 939
\bibitem{}
Yi I., Vishniac E. T., 1999, ApJL, 516, 87

\end{thebibliography}
\end{document}